\documentstyle[12pt,subeqn,axodraw]{article}

\newcommand\vek[1]{\mbox{\rmfamily\bfseries\itshape#1}}

\def\rd{{\rm d}}

\begin{document}

\title{Thermodynamic Properties of the One-Dimensional
Two-Component Log-Gas}
\author{L. {\v S}amaj$^{1,2}$}
\maketitle
\begin{abstract}
We consider a one-dimensional continuum gas of pointlike positive 
and negative unit charges interacting via a logarithmic potential.
The mapping onto a two-dimensional boundary sine-Gordon field theory 
with zero bulk mass provides the full thermodynamics (density-fugacity
relationship, specific heat, etc.) of the log-gas
in the whole stability range of inverse temperatures $\beta<1$.
An exact formula for the excess chemical potential of a ``foreign'' 
particle of an arbitrary charge, put into the log-gas, is derived.
The results are checked by a small-$\beta$ expansion and
at the collapse $\beta=1$ point.
The possibility to go beyond the collapse temperature is discussed.
\end{abstract}

\noindent {\bf KEY WORDS:} Two-component plasma; one dimension;
logarithmic interaction; thermodynamics; boundary sine-Gordon model; 
integrability.

\medskip
\noindent LPT Orsay 01-09

\vfill

\noindent
$^1$ Laboratoire de Physique Th\'eorique, Universit\'e de Paris-Sud,
B\^atiment 210, 91405 Orsay Cedex, France (Unit\'e Mixte de
Recherche no. 8627 - CNRS); e-mail: samaj@th.u-psud.fr

\noindent 
$^2$ On leave from the Institute of Physics, Slovak Academy of 
Sciences, Bratislava, Slovakia

\newpage

\renewcommand{\theequation}{1.\arabic{equation}}
\setcounter{equation}{0}

\section{Introduction}
The model under consideration is a two-dimensional (2D) two-component (TC)
classical Coulomb gas confined to a 1D manifold, the circle of radius $R$
or its $R\to \infty$ limit -- the straight line.
It is usually named as the symmetric 1D TC log-gas.
The model consists of a mixture of mobile pointlike 
particles $\{ j \}$ of positive and negative, say unit, charges
$\{ q_j = \pm \}$, localized at continuous angle positions 
$\{ \varphi_j \}$ on the circle of length $L=2 \pi R$.
The interaction energy reads
\begin{equation} \label{1.1}
E(\{ q_j, \varphi_j \}) = \sum_{j<k} q_j q_k 
v(\varphi_j,\varphi_k)
\end{equation}
where $v(\varphi,\varphi')$ is the 2D Coulomb
potential with imposed periodic boundary conditions,
\begin{equation} \label{1.2}
v(\varphi,\varphi') = - \ln \left\{ \left( {R\over r_0}
\right) \vert {\rm e}^{{\rm i}\varphi}-{\rm e}^{{\rm i}\varphi'}
\vert \right\}
\end{equation}
The length constant $r_0$, which fixes the zero point of energy,
will be set for simplicity to unity.
In the thermodynamic limit $R\to\infty$, which is of interest
in this paper, introducing the straight line position variable 
$x\in (-\infty,\infty)$ via $x = \varphi R$, 
the interaction potential (\ref{1.2}) 
takes the familiar logarithmic form
\begin{equation} \label{1.3}
v(x,x') = - \ln \left( \vert x - x' \vert \right)
\end{equation}

For pointlike particles, the interaction Boltzmann factor of a 
positive-negative pair of charges at distance $x$, $x^{-\beta}$
where $\beta$ is the inverse temperature, is integrable on 
a 1D manifold at small $x$ if and only if $\beta<1$ 
(at large $x$, there is no problem because the interaction is
screened by the system), with $\beta = 1$ being the collapse point.
In its conductive phase, the model exhibits poor screening properties, 
a typical feature of a $D$-dimensional Coulomb system confined to a
domain of dimension $D-1$ \cite{Forrester1}.
As a consequence, the leading non-oscillatory term of the large-distance 
asymptotic decay of the charge-charge correlation is algebraically
slow $\sim -1/[\beta (\pi x)^2]$ \cite{Alastuey}.

The 1D TC log-gas, with $\pm$ charges required to alternate in space,
is equivalent to the impurity Kondo problem 
\cite{Anderson1} - \cite{Schotte}.

The 1D TC log-gas, without any restriction on the charge order,
is related to the problem of quantum Brownian motion of a particle in 
a 1D periodic potential \cite{Schmid} - \cite{Guinea}.
The model was also studied in connection with 
the problem of non-equilibrium quantum transport through 
a point contact in a 1D Luttinger liquid \cite{Kane} - \cite{Fendley1}, 
having a realization in resonant tunneling-transport 
experiments between edge states in fractional quantum Hall effect 
devices \cite{Milliken}.

The 1D TC log-gas with a hard core, or with some other short-distance
regularization of the Coulomb interaction potential, 
undergoes a 1D counterpart of the Kosterlitz-Thouless
transition from a conducting phase to a low-temperature insulating 
phase around $\beta=2$.
An approximate analytic study of the gas with $\log(1+\vert x\vert)$ 
interaction was given in ref. \cite{Schultz}.
The lattice version of the model was exactly solved at $\beta = 1, 2, 4$ 
and its conductor-insulator phase diagram was conjectured in refs.
\cite{Forrester2} - \cite{Forrester4}.

Recently \cite{Samaj}, the bulk thermodynamic properties
(density-fugacity relationship, free energy, internal energy,
specific heat, etc.) of the infinite 2D TC plasma have 
been obtained exactly, in the whole model's stability range of
temperatures, by mapping the plasma onto a 2D bulk sine-Gordon field 
theory and then using recent results about that integrable field theory.
The aim of this paper is to derive the full thermodynamics of
the confined plasma, the 1D TC log-gas, via its relationship 
to the 2D boundary sine-Gordon field theory with zero bulk mass.
The particle nature of the statistical model permits one to
check the obtained behaviour of thermodynamic quantities close 
to the collapse $\beta=1$ point, and to suggest a possible
analytic extension of the results beyond that collapse point.

The mapping onto the 2D boundary sine-Gordon model
is outlined within the grandcanonical formalism in section 2.
To give to the mapping a precise meaning, we analyse the 
short-distance behaviour of the pair distribution function
for two basic combinations of particle charges.
The relationship between the ordinary statistical mechanics
\cite{Jancovici1}, \cite{Hansen} and the Conformal Perturbation
theory \cite{Fateev1} is established.
As a byproduct, for an ``external'' (or, probably more adequately,
``foreign'') particle with an arbitrary charge put into
the log-gas, a relation between its excess chemical potential
and the one-point expectations of the exponential boundary
field in the sine-Gordon model is derived.

Section 3, devoted to the derivation of full thermodynamics 
of the 1D TC log-gas, is based on an exact fromula for 
the above one-point expectations \cite{Fateev2}, 
obtained by using a ``reflection'' relationship
between the 2D boundary Liouville and sinh-Gordon theories.

Section 4 brings a discussion about a possible analytic extension 
of the acquired results to the collapse region $1\le \beta \le 2$.
In this region, one could attach to the particles a small hard core 
$\sigma$ in order to prevent the collapse, then calculate 
thermodynamic quantities, and at the end take
the $\sigma \to 0$ limit.
In this limit, while the free energy and the internal energy per
particle diverge, the specific heat, truncated correlation functions,
etc., are expected to remain finite \cite{Hauge}.
An analytic continuation of the formula for the specific heat
(at constant volume) predicts an infinite sequence of
phase transitions from the conducting plasma phase $\beta \le 1$ 
to the insulator region $\beta \ge 2$.
In two dimensions, such a phenomenon was predicted in ref.
\cite{Gallavotti} but later denied in ref. \cite{Fisher}.

Whenever possible, the results are checked by a systematic
small-$\beta$ (high-temperature) expansion, using a renormalized 
Mayer expansion technique developed in refs. \cite{Samaj},
\cite{Deutsch}, \cite{Jancovici2} (see Appendix), and close to the 
collapse $\beta = 1$ point, by applying arguments in the spirit of an
independent-pair approximation \cite{Hauge}, which assumes
a dominant contribution from almost collapsed positive-negative
pairs of charges to the configurational integral.

\renewcommand{\theequation}{2.\arabic{equation}}
\setcounter{equation}{0}

\section{Mapping onto the boundary sine-Gordon theory}
Let us first examine a general TC plasma of $q=\pm 1$ charges in an
infinite 2D space of points ${\vek r} = (x,y)$.
We will work in the grand canonical ensemble, with position-dependent
fugacities $z_+({\vek r})$ and $z_-({\vek r})$ of the positive
and negative particles, respectively.
In infinite space, $-\Delta/(2\pi)$ is the inverse operator
of the Coulomb potential $-\ln(\vert {\vek r}\vert)$.
Using the standard procedure (see, e.g., ref. \cite{Minnhagen}),
the grand partition function $\Xi$ of the plasma at inverse
temperature $\beta$, considered as the
functional of particle fugacities, can be turned into
\begin{equation} \label{2.1}
\Xi[z_+,z_-] = {\int {\cal D}\phi \exp \left[ \int {\rm d}^2 r \left(
{1\over 4\pi} \phi \Delta \phi + 
z_+({\vek r}) {\rm e}^{{\rm i}\sqrt{\beta}\phi} +
z_-({\vek r}) {\rm e}^{-{\rm i}\sqrt{\beta}\phi} \right) \right]
\over \int {\cal D}\phi \exp \left( \int {\rm d}^2 r 
{1\over 4\pi} \phi \Delta \phi \right)}
\end{equation}
Here, $\phi({\vek r})$ is a real scalar field, $\int {\cal D} \phi$ 
denotes the functional integration over this field and the fugacities
are renormalized by a self-energy term.
The consideration of
\begin{equation} \label{2.2}
z_{q}({\vek r}) = z_{q}(x) \delta(y)
\end{equation}
confines the charges to the $x$-line,
\begin{equation} \label{2.3}
\Xi[z_+,z_-] = {\int {\cal D}\phi \exp \left\{ \int {\rm d}^2 r 
\left( {1\over 4\pi} \phi \Delta \phi \right) + \int_{-\infty}^{\infty} 
{\rm d} x \left[ z_+(x) {\rm e}^{{\rm i}\sqrt{\beta}\phi_B} +
z_-(x) {\rm e}^{-{\rm i}\sqrt{\beta}\phi_B} \right] \right\}
\over \int {\cal D}\phi \exp \left( \int {\rm d}^2 r 
{1\over 4\pi} \phi \Delta \phi \right)}
\end{equation}
where $\phi_B(x) \equiv \phi(x,y=0)$ is the boundary field.

To reformulate the field theory (\ref{2.3}) as a boundary problem, 
in formal analogy with ref. \cite{Callan},
one introduces two new fields
\begin{subequations} \label{2.4}
\begin{eqnarray}
\phi_e(x,y) & = & {1\over \sqrt{2}} \left[ \phi(x,y) + \phi(x,-y)
\right] \label{2.4a} \\
\phi_o(x,y) & = & {1\over \sqrt{2}} \left[ \phi(x,y) - \phi(x,-y)
\right] \label{2.4b}
\end{eqnarray}
\end{subequations}
defined only in the upper half-plane $y\ge 0$.
The even field has a Neumann boundary condition 
$\partial_y \phi_e \vert_{y=0} = 0$ and the odd field has a
Dirichlet boundary condition $\phi_o \vert_{y=0} = 0$.
It holds
\begin{equation} \label{2.5}
\int {\rm d}^2 r \phi \Delta \phi = 
\int_{y>0} {\rm d}^2 r (\phi_e \Delta \phi_e + \phi_o \Delta \phi_o)
\end{equation}
The odd field, contributing only by its free-field part $\phi_o
\Delta \phi_o$, disapears from (\ref{2.3}) by numerator-denominator
cancelation.
By integration per partes, the term $\phi_e \Delta \phi_e$ can be
rewritten as $-(\nabla \phi_e)^2$, with a vanishing contribution 
from the boundary.
Considering that $\phi_B(x) = \phi_e(x,y=0)/\sqrt{2}$, and 
renaming then $\phi_e$ as $\phi$, (\ref{2.3}) transforms to
\begin{equation} \label{2.6}
\Xi[z_+,z_-] = {\int {\cal D} \phi \exp \left( -S_{sG}[z_+,z_-]\right)
\over \int {\cal D} \phi \exp \left( -S_{sG}[0,0]\right) }
\end{equation}
with the action
\begin{subequations} \label{2.7}
\begin{eqnarray} 
S_{sG}[z_+,z_-] & = & \int_{-\infty}^{\infty} {\rm d}x
\int_0^{\infty} {\rm d}y {1\over 4\pi} (\nabla \phi)^2 -
\int_{-\infty}^{\infty} {\rm d}x \left[ z_+(x) 
{\rm e}^{{\rm i}b\phi_B} + z_-(x) {\rm e}^{-{\rm i}b\phi_B} \right] 
\label{2.7a} \\
\beta & = & 2 b^2 \label{2.7b}
\end{eqnarray}
\end{subequations}
where again $\phi_B(x) \equiv \phi(x,y=0)$ and the $\phi$-field
has the Neumann boundary condition.
For uniform and equivalent charge fugacities, $z_+(x) = z_-(x) = z$,
one gets the boundary sine-Gordon model with zero bulk mass, 
\begin{equation} \label{2.8}
S_{sG}(z)  =  \int_{-\infty}^{\infty} {\rm d}x
\int_0^{\infty} {\rm d}y {1\over 4\pi} (\nabla \phi)^2 - 2 z
\int_{-\infty}^{\infty} {\rm d}x \cos(b\phi_B)
\end{equation}

The sine-Gordon representation of the multi-particle densities can be 
obtained from the functional generator (\ref{2.6}), (\ref{2.7}) 
in a straightforward way.
The density of particles of one sign is
\begin{equation} \label{2.9}
n_{q}  =  z_{q} {\delta \ln \Xi \over \delta
z_{q}(x) } \big\vert_{z_{q}(x)=z} 
=  z_{q} \langle {\rm e}^{{\rm i}q b \phi_B}
\rangle_{sG} 
\end{equation}
where $\langle \cdots \rangle_{sG}$ denotes the averaging 
over the sine-Gordon action (\ref{2.8}).
Here, although the charge symmetry is considered, 
i.e., $n_+ = n_- = n/2$ ($n$ is the total number density of particles), 
we leave the $q$-subscript in order to make transparent identities like
$\langle {\rm e}^{{\rm i}b\phi_B} \rangle_{sG} =
\langle {\rm e}^{-{\rm i}b\phi_B} \rangle_{sG}$.
This identity is a special case of a general symmetry relation
\begin{equation} \label{2.10}
\langle {\rm e}^{{\rm i}a \phi_B} \rangle_{sG} =
\langle {\rm e}^{-{\rm i}a\phi_B} \rangle_{sG} ,
\quad \quad \quad a\ {\rm arbitrary}
\end{equation}
which results from the invariance of the sine-Gordon action (\ref{2.8})
with respect to the transformation $\phi \to -\phi$.
The pair distribution function $g_{q,q'}(x,x')$
is given by
\begin{eqnarray} \label{2.11}
g_{q,q'}(\vert x-x' \vert) & = & 
\left( {z_{q}\over n_{q}} \right)
\left( {z_{q'}\over n_{q'}} \right) 
{1\over \Xi} {\delta^2 \Xi \over 
\delta z_{q}(x) \delta z_{q'}(x') } 
\big\vert_{z_{q}(x)=z} \nonumber \\
& = & \left( {z_{q}\over n_{q}} \right) 
\left( {z_{q'}\over n_{q'}} \right) 
\langle {\rm e}^{{\rm i}q b \phi_B(x)} 
{\rm e}^{{\rm i}q' b \phi_B(x')} \rangle_{sG} 
\end{eqnarray}
and so on.

In statistical mechanics, 
the short-distance behaviour of the pair distribution
function $g_{q,q'}(\vert x-x' \vert)$
is dominated by the Boltzmann factor of the pair potential,
\begin{equation} \label{2.12}
g_{q,q'}(x,x') \sim C_{q,q'} 
\vert x-x' \vert^{\beta q q'} \quad \quad \quad
{\rm as}\ \vert x-x' \vert \to 0
\end{equation}
(provided $\beta$ is small enough; see below).
The prefactors $C_{q,q'}$ are related to a free
energy difference \cite{Jancovici1}, \cite{Hansen}.
For $q' = -q$, one has
\begin{equation} \label{2.13}
C_{q,-q} = \exp \left[ \beta \left( \mu_+^{{\rm ex}}
+ \mu_-^{{\rm ex}} \right) \right]
\end{equation}
where $\mu_{q}^{{\rm ex}}$ is the excess chemical potential
of species $q$.
Strictly speaking, relations (\ref{2.12}) and (\ref{2.13})
are valid only for $\beta<1$; at $\beta=1$, due to the
collapse of positive-negative pairs of charges, 
$\mu_{\pm}^{{\rm ex}} \to \infty$.
On the other hand, for the truncated two-body densities
$n_{q,q'}^{({\rm T})}(x,x') = n_{q} n_{q'} [g_{q,q'}(x,x')-1]$,
the short-distance asymptotics analogous to that given by
eq. (\ref{2.12}) with $q'=-q$ takes place also for $\beta>1$. 
For $q' = q$, one has
\begin{equation} \label{2.14}
C_{q,q} = \exp \left[ \beta \left( 2 \mu_{q}^{{\rm ex}}
- \mu_{2q}^{{\rm ex}} \right) \right]
\end{equation}
Here, we use an extended definition of the excess chemical potential 
for ``foreign'' ions with arbitrary charges, put into the underlying 
electrolyte: 
$\mu_Q^{{\rm ex}}$ = reversible work which has to be done in order 
to bring a foreign particle of charge $Q$ from infinity into 
the bulk interior of the 1D TC plasma of unit $\pm$ charges 
(the consequent breaking of the system neutrality
has a negligible effect in the thermodynamic limit).
This quantity is of evident importance in chemistry.
The interaction Boltzmann factor of the $Q$ charge and an
opposite unit charge at distance $x$, $x^{-\beta \vert Q \vert }$, 
is integrable at small $x$ if and only if $\vert Q \vert \beta<1$.
The stability region for $\mu_Q^{{\rm ex}}$ therefore is expected
to be $\beta< 1/\vert Q\vert$ with $\mu_Q^{{\rm ex}}$ diverging
just at $\beta = 1/\vert Q \vert$.
As a consequence, relations (\ref{2.12}) and (\ref{2.14}) are
valid only for $\beta < 1/2$. 
For $\beta > 1/2$, the two-point correlator 
$\propto \vert x-x' \vert^{1-\beta}$; the analysis of this change
in the short-distance expansion goes beyond the scope of 
the present work. 
With regard to the thermodynamic relation
\begin{equation} \label{2.15}
\ln \left( {z_{q} \over n_{q}} \right) = \beta
\mu_{q}^{{\rm ex}}
\end{equation}
we conclude that, for $q = \pm$, it holds
\begin{equation} \label{2.16}
\langle {\rm e}^{{\rm i}q b \phi_B} \rangle_{sG}
= \exp \left( - \beta \mu_{q}^{{\rm ex}} \right)
\end{equation}
and
\begin{subequations} \label{2.17}
\begin{eqnarray}
\langle {\rm e}^{{\rm i}q b \phi_B(x)} 
{\rm e}^{-{\rm i}q b \phi_B(x')} \rangle_{sG} 
& \sim & \vert x-x' \vert^{-2b^2} \quad \quad \quad
{\rm as}\ \vert x - x' \vert \to 0 \label{2.17a} \\
\langle {\rm e}^{{\rm i}q b \phi_B(x)} 
{\rm e}^{{\rm i}q b \phi_B(x')} \rangle_{sG} 
& \sim & {\rm e}^{-\beta \mu_{2q}^{{\rm ex}}}
\vert x-x' \vert^{2b^2} \quad \quad \quad
{\rm as}\ \vert x - x' \vert \to 0 \label{2.17b}
\end{eqnarray}
\end{subequations}
where we have used the equality $\beta = 2b^2$, see
formula (\ref{2.7b}).

In quantum field theory, the sine-Gordon model (\ref{2.8})
can be regarded as a conformal field theory perturbed
by the boundary cos-field.
The short distance expansions for multi-point correlation 
functions are then obtainable by using
the Operator Product expansions (OPE) \cite{Wilson}.
The vacuum one-point expectations, in our case
$\langle \exp({\rm i}a\phi_B)\rangle_{sG}$ with an arbitrary 
value of $a$, are the basic objects of the OPE scheme which contain 
the whole nonperturbative information about the system.
For the product of two primary fields
${\rm e}^{{\rm i}a_1\phi}(x_1){\rm e}^{{\rm i}a_2\phi}(x_2)$,
the OPE has the form
\begin{equation} \label{2.18}
{\rm e}^{{\rm i}a_1\phi_B}(x_1){\rm e}^{{\rm i}a_2\phi_B}(x_2)
= \sum_{n=-\infty}^{\infty} \left[ C_{a_1,a_2}^{n,0}(\vert x_1
- x_2 \vert) {\rm e}^{{\rm i}(a_1+a_2+n b)\phi_B}(x_2)
+ \cdots \right]
\end{equation}
where the coefficients $C$ are computable within the Conformal
Perturbation theory \cite{Fateev1}.
By successive application of (\ref{2.18}) the short-distance
behaviour of any multi-point correlation function of the exponential 
field can be reduced to one-point functions.
For the present model, the leading short-distance term of
the two-point function reads
\begin{equation} \label{2.19}
\langle {\rm e}^{{\rm i}q b \phi_B(x)} 
{\rm e}^{{\rm i}q' b \phi_B(x')} \rangle_{sG} 
\sim \langle {\rm e}^{{\rm i}(q+q') b \phi_B} \rangle_{sG} 
\vert x-x' \vert^{2q q' b^2} \quad \quad \quad
{\rm as}\ \vert x - x' \vert \to 0
\end{equation}
For $q' = -q$, formula (\ref{2.17a}) is reproduced.
For $q' = q$, the combination of (\ref{2.17b}) and (\ref{2.19}) 
implies the relationship
\begin{equation} \label{2.20}
\exp \left( -\beta \mu_{2q}^{{\rm ex}} \right)
= \langle {\rm e}^{{\rm i}2q b \phi_B} \rangle_{sG} 
\end{equation}

The above formalism generalizes straightforwardly to $Q$-particle
distribution functions ($Q$ positive integer), with the result
\begin{equation} \label{2.21}
\exp \left( -\beta \mu_Q^{{\rm ex}} \right)
= \langle {\rm e}^{{\rm i}Q b \phi_B} \rangle_{sG} 
\end{equation}
Eqs. (\ref{2.16}) and (\ref{2.20}) are the lowest $Q=1$ and $Q=2$
cases, respectively, of this formula.
We expect the validity of relation (\ref{2.21}) also for
noninteger values of $Q$.
An analogous formula can be derived for the infinite 
(unconfined) 2D TC plasma.

\renewcommand{\theequation}{3.\arabic{equation}}
\setcounter{equation}{0}

\section{Thermodynamics}
The 2D boundary sine-Gordon theory with the action (\ref{2.8})
has a discrete symmetry $\phi \to \phi + 2\pi n/b$ with
any integer $n$.
In the domain $0<b^2<1$ this symmetry is spontaneously broken
and the theory has infinitely many ground states $\vert 0_n \rangle$,
characterized by the associate expectation values of the field $\phi$,
$\langle \phi \rangle_n = 2\pi n/b$.
One has to choose one of these ground states, say $\vert 0_0 \rangle$,
and consider $\langle \cdots \rangle_{sG} 
\equiv \langle \cdots \rangle_0$.
The parameter $z$, which renormalizes multiplicatively, gets a precise 
meaning when one fixes the normalization of the field $\cos(b \phi_B)$.
The conformal normalization \cite{Fateev2}, based on the OPE scheme
(\ref{2.18}) and consistent with formula (\ref{2.17a}) derived
via the 1D TC log-gas, corresponds to the short-distance limit 
of the two-point function
\begin{equation} \label{3.1}
\langle {\rm e}^{{\rm i}a \phi_B(x)} 
{\rm e}^{-{\rm i}a \phi_B(x')} \rangle_{sG} 
\sim \vert x-x' \vert^{-2a^2} \quad \quad \quad
{\rm as}\ \vert x - x' \vert \to 0
\end{equation}
Under normalization (\ref{3.1}), the expectation value of the
exponential boundary field was obtained in ref. \cite{Fateev2},
\begin{eqnarray} \label{3.2}
\langle {\rm e}^{{\rm i}a\phi_B} \rangle_{sG} & = &
\left[ {2^{b^2} \pi z \over \Gamma(b^2)} \right]^{a^2/(1-b^2)} 
\exp \left\{ \int_0^{\infty} {{\rm d}t \over t} \left[
{\left( {\rm e}^t - 1 + {\rm e}^{t(1-b^2)} + {\rm e}^{-b^2 t} \right)
\sinh^2(abt) \over 2 \sinh(b^2t) \sinh(t) \sinh((1-b^2)t)}
\right. \right. \nonumber \\ & & \left. \left. 
\hskip 6truecm
- a^2 \left( {1\over \sinh((1-b^2)t)} + {\rm e}^{-t} \right)
\right] \right\}
\end{eqnarray} 
where $\vert {\rm Re} ~ 2a \vert < 1/b$.
This result was derived under assumption of a ``reflection''
relationship between the 2D boundary Liouville and sinh-Gordon theories
(with zero bulk mass), and the consequent analytic continuation
of the latter theory in the $b$-parameter to the boundary 
sine-Gordon model.
For the case of special interest $a=b$, using the integral 
representation of the Gamma function
\begin{equation} \label{3.3}
\ln \Gamma(x) = \int_0^{\infty} {{\rm d}t \over t} {\rm e}^{-t}
\left[ x-1+{{\rm e}^{-(x-1)t} - 1 \over 1- {\rm e}^{-t}} \right],
\quad \quad \quad {\rm Re}~ x >0
\end{equation}
after some algebra, relation (\ref{3.2}) takes a simpler form
\begin{equation} \label{3.4}
\langle {\rm e}^{{\rm i}b\phi_B} \rangle_{sG} =
{1\over 4 \pi^{3/2} (1-b^2) z} 
\Gamma \left( {1-2 b^2 \over 2-2 b^2} \right)
\Gamma \left( {b^2 \over 2-2b^2} \right)
\left[ {2\pi z \over \Gamma(b^2)} \right]^{1/(1-b^2)}
\end{equation}
On account of the symmetry relation (\ref{2.10}), it holds
\begin{equation} \label{3.5}
\langle {\rm e}^{{\rm i}b\phi_B} \rangle_{sG} =
{1\over 2} {\partial \over \partial z} \lim_{L\to\infty}
{1\over L} \ln \Xi_L
\end{equation}
where $L$ is the length of the system.
Since $\Xi_L(z=0) = 1$, one has
\begin{equation} \label{3.6}
\lim_{L\to\infty} {1\over L} \ln \Xi_L =
{1\over 2 \pi^{3/2}} 
\Gamma \left( {1-2 b^2 \over 2-2 b^2} \right)
\Gamma \left( {b^2 \over 2-2b^2} \right)
\left[ {2\pi z \over \Gamma(b^2)} \right]^{1/(1-b^2)}
\end{equation}

Relating the sine-Gordon parameters with those of the TC log-gas,
$\beta = 2b^2$ from (\ref{2.7b}) and $n/(2z) = \langle 
\exp({\rm i}b\phi_B)\rangle_{sG}$ according to (\ref{2.9}),
formulae in the above paragraph imply the explicit $n-z$
relation for the plasma:
\begin{equation} \label{3.7}
{n^{1-\beta/2}\over z} = {2 \pi \over \Gamma(\beta/2)}
\left[ {1\over 2 \pi^{3/2}(1-\beta/2)} 
\Gamma\left( {1-\beta \over 2-\beta} \right)
\Gamma\left( {\beta/2 \over 2-\beta} \right) \right]^{1-\beta/2}
\end{equation}
The small-$\beta$ expansion of the rhs of (\ref{3.7}) reads
\begin{eqnarray} \label{3.8}
{n^{1-\beta/2}\over z} & = & 2 \beta^{\beta/2} \exp \big\{
(C+\ln \pi) {\beta\over 2} + {1\over 12} \left[ \pi^2
+7 \zeta(3) \right] \left( {\beta\over 2} \right)^3  \nonumber \\
& & + {1\over 12} \left[ \pi^2 + 6 \zeta(3) \right] 
\left( {\beta\over 2} \right)^4  + O(\beta^5) \big\}
\end{eqnarray}
where $C$ is the Euler's constant and $\zeta$ is the Riemann's
zeta function.
Formula (\ref{3.8}) is checked in the Appendix by using 
a renormalized Mayer expansion in density.
For fixed $z$, relation (\ref{3.7}) tells us that the particle
density $n$ exhibits the expected collapse singularity
as $\beta \to 1^-$:
\begin{equation} \label{3.9}
n \sim {4 z^2 \over 1-\beta }
\end{equation}
The form of this singularity can be reproduced indirectly 
by using a perfect-screening sum rule for the one-body densities,
\begin{equation} \label{3.10}
n_{q} = \int {\rm d} x \left[ 
n_{q,-q}^{{(\rm T)}}(x) -
n_{q,q}^{{(\rm T)}}(x) \right]
\end{equation}
where the truncated two-body densities
$n_{q,q'}^{({\rm T})}(x,x') = n_{q} n_{q'}
[g_{q,q'}(x,x')-1]$.
For $\beta\to 1^-$, the integral in (\ref{3.10}) is dominated
by the short-distance behaviour of 
$n_{q,-q}^{({\rm T})}(x) \sim z^2 \vert x \vert^{-\beta}$
[see relations (\ref{2.12}), (\ref{2.13}) and (\ref{2.15})].
Then,
\begin{equation} \label{3.11}
{n\over 2} = \int_{-\lambda}^{\lambda} {\rm d}x {z^2 \over
\vert x \vert^{\beta}} = {2 z^2 \over 1-\beta} \lambda^{1-\beta}
= {2 z^2 \over 1-\beta} + O(1)
\end{equation}
where $\lambda$ is a length over which the Coulomb interaction
is screened by the system, and the singular behaviour (\ref{3.9})
is justified.

To get the full thermodynamics of the 1D TC log-gas, we pass from 
the grandcanonical to the canonical ensemble via 
the Legendre transformation
\begin{equation} \label{3.12}
F_L(\beta,N) = \Omega_L + \mu N
\end{equation}
Here, the total particle number $N = n L$, the chemical potential
\begin{equation} \label{3.13}
\mu(\beta,n) = {1\over \beta} \ln z(\beta,n)
\end{equation}
and the grand potential $\Omega$ is defined by
\begin{equation} \label{3.14}
- \beta \Omega_L = \ln \Xi_L = L \left( 1 - {\beta \over 2}
\right) n
\end{equation}
where we have combined eqs. (\ref{3.4}) and (\ref{3.6}).
Note that the thermodynamic relation $\ln \Xi_L = \beta P L$
($P$ is the pressure) implies the well-known equation of state
\begin{equation} \label{3.15}
\beta P = \left( 1 - {\beta \over 2} \right) n
\end{equation}
The dimensionless specific free energy $f$,
$f = \beta F_L / N$, is then written as
\begin{eqnarray} \label{3.16}
f(\beta,n) & = &  \left( 1- {\beta \over 2} \right) \left[
-1 + \ln n + \ln \left( 1- {\beta \over 2} \right) \right]
-{\beta \over 2} \ln 2 
+ {1\over 2} \left( 1- {3 \beta \over 2} \right) \ln \pi
\nonumber \\ & & 
+ \ln \Gamma\left({\beta \over 2}\right) - 
\left( 1- {\beta \over 2} \right) \left[
\ln \Gamma \left( {1-\beta \over 2-\beta} \right) +
\ln \Gamma \left( {\beta/2 \over 2-\beta} \right) \right]
\end{eqnarray}
According to the elementary thermodynamics, the (excess)
internal energy per particle, $u^{{\rm ex}} = \langle E \rangle$,
and the (excess) specific heat at constant volume per particle,
$c_V^{{\rm ex}} = C_V/N$, are given by
\begin{subequations}
\begin{eqnarray}
u^{{\rm ex}} & = & {\partial \over \partial \beta} f(\beta,n)
\label{3.17a} \\
{c_V^{{\rm ex}} \over k_B} & = & - \beta^2
{\partial^2 \over \partial \beta^2} f(\beta,n) \label{3.17b}
\end{eqnarray}
\end{subequations}
For $c_V^{{\rm ex}}$, one gets explicitly
\begin{eqnarray} \label{3.18}
{c_V^{{\rm ex}} \over k_B}  & = & {\beta^2 \over 4 (2-\beta)^3}
\big\{ 2 \psi' \left( {1-\beta \over 2-\beta} \right)
+ (\beta-2)^3 \psi'\left( {\beta\over 2} \right)
\nonumber \\ & & 
- 2 (\beta-2)^2 + 2 \psi'\left( {\beta/2 \over 2-\beta} \right)
\big\}
\end{eqnarray}
where $\psi(x) = {\rm d} [\ln \Gamma(x)]/{\rm d}x$ is 
the psi function.
The series representation of its first derivative reads
\begin{equation} \label{3.19}
\psi'(x) = \sum_{j=0}^{\infty} {1\over (x+j)^2}
\end{equation}
The expansion of $c_V^{{\rm ex}}/k_B$ around the collapse
point $\beta\to 1^-$ results in the Laurent series
\begin{equation} \label{3.20}
{c_V^{{\rm ex}} \over k_B} = {1\over 2 (1-\beta)^2} -
{3\over 2 (1-\beta)} + \left( {3\over 2} + {5 \pi^2 \over 24} \right)
+ O(1-\beta)
\end{equation}
The leading term of this series can be reproduced by regarding
that, under assumption (\ref{3.11}),
\begin{equation} \label{3.21}
f(\beta,n) \sim {1\over 2} \ln (1-\beta) - {1\over 2} ( 1- \beta )
\ln \lambda
\quad \quad
{\rm as}\ \beta\to 1^-
\end{equation}
and then applying (\ref{3.17b}).

As concerns the excess chemical potential of a foreign
particle of charge $Q$ in the plasma, setting $a=b Q$ 
and $b^2=\beta/2$ in (\ref{3.2}), relation (\ref{2.21}) gives
\begin{eqnarray} \label{3.22}
-\beta \mu_Q^{{\rm ex}} & = & {\beta Q^2 \over 2-\beta}
\ln \left[ {2^{\beta/2} \pi z \over \Gamma(\beta/2)} \right]
+ \int_0^{\infty} {{\rm d}t \over t} \left[
{\left( {\rm e}^t - 1 + {\rm e}^{t(1-\beta/2)} + 
{\rm e}^{-\beta t/2} \right)
\sinh^2(Q\beta t/2) \over 2 \sinh(\beta t/2) \sinh(t) 
\sinh((1-\beta/2)t)} \right.
\nonumber \\ & & \hskip 6truecm \left. 
- {\beta Q^2 \over 2} \left( 
{1\over \sinh((1-\beta/2)t)} + {\rm e}^{-t} \right) \right]
\end{eqnarray}
In the limit of large $t$, the first integrand behaves like 
$(1/t)\exp[(\beta \vert Q \vert -1)t]$, and therefore the integral
is finite if $\beta < 1/\vert Q \vert$.
As was argued in the previous section, this is the true stability 
region for $\mu_Q^{{\rm ex}}$.
For $\vert Q \vert <1$, $\mu_Q^{{\rm ex}}$ remains finite also in 
a subspace of the collapse region, namely 
in the interval $1 < \beta <1/\vert Q \vert$.
Everything what was said above holds only when $\beta<2$. 
Approaching the conductor-insulator phase transition at 
point $\beta=2$, the first term on the rhs of (\ref{3.22}) 
diverges for an arbitrary $Q$.

\renewcommand{\theequation}{4.\arabic{equation}}
\setcounter{equation}{0}

\section{Discussion}
All results in this paper were derived
under the assumption that the system is stable against collapse,
i.e., in the range of inverse temperatures $\beta<1$.
The only exception is represented by the excess chemical
potential $\mu^{{\rm ex}}_Q$ of a ``foreign'' $Q$-charged
particle embedded in the plasma: 
if $\vert Q \vert <1$, relation (\ref{3.22}) applies 
also to the collapse region, up to $\beta = 1/\vert Q\vert$.
A natural question arises whether there is a possibility to
make an analytic continuation of the explicit result for some 
bulk quantity into the collapse region $1<\beta<2$.
The best candidate for such a continuation is the specific heat,
formula (\ref{3.18}), which might be in the sense explained in the
Introduction a well defined finite quantity also in the
collapse region.

Before going further, let us recall the work of Gallavotti and
Nicol{\'o} \cite{Gallavotti} concerning the infinite 
2D TC Coulomb gas of unit $\pm$ charges.
In two dimensions, the particle collapse occurs at $\beta=2$ and 
the Kosterlitz-Thouless phase transition takes place around $\beta=4$.
By studying the Mayer series of the specific grand potential in 
{\it fugacity} it was proven in ref. \cite{Gallavotti} that each
term of the series converges in the insulator region $\beta>4$.
For $\beta\le 4$, the existence of infinitely many thresholds
\begin{equation} \label{4.1}
2D: \quad \quad \quad \beta_N = 4 \left( 1 - {1\over 2N} \right),
\quad \quad \quad N = 1, 2, \ldots
\end{equation}
was observed: only the Mayer series' coefficients up to order $2N$ 
are finite if $\beta > \beta_N$.
Points $\{ \beta_N \}$ were conjectured to correspond to a sequence 
of transitions from the pure multipole insulating phase ($\beta>4$)
to the conducting plasma phase ($\beta <2$) via an infinite number 
of intermediate phases.
Although the mathematics used in \cite{Gallavotti} was quite
complicated, there exists a simple explanation of the above
findings \cite{Jancovici3}.
For a neutral choice of $N$ positive and $N$ negative
charges in a disk of radius $R$, the most relevant contribution
to the coefficient of the $z^{2N}$ term comes from the complete-star
integral, after approximating the Mayer function by $\exp(-\beta v)$,
\begin{equation} \label{4.2}
{1\over \pi R^2} \int_{\sigma}^R \prod_{i=1}^N 
{\rm d}^2 r_i {\rm d}^2 r'_i ~
{\prod_{i<j} \vert {\vek r}_i - {\vek r}_j \vert^{\beta}
\vert {\vek r}'_i - {\vek r}'_j \vert^{\beta}
\over \prod_{i,j} \vert {\vek r}_i - {\vek r}'_j \vert^{\beta}}
\end{equation}
where $\sigma$ is a small hard-core diameter and ${\vek r}$ (${\vek r}'$)
denote spatial coordinates of positive (negative) charges.
Rescaling ${\vek r}_i \to R {\vek s}_i$ and
${\vek r}'_i \to R {\vek s}'_i$, one gets
$(4.2) = R^{N(4-\beta)-2} \times f(\sigma/R)$ which diverges as
$R\to\infty$ just when $\beta<\beta_N$.
In two dimensions, the formula for $c_V^{{\rm ex}}/k_B$ in the
region $\beta<2$ is presented in ref. \cite{Samaj}, relation (56).
The only source of the singularities of $c_V^{{\rm ex}}/k_B$,
extended into the region $2<\beta<4$, is the term $\propto
\sin^{-2}(\pi\beta/(4-\beta))$, which gives a double pole at
\begin{equation} \label{4.3}
{\bar \beta}_N = 4 \left( 1 - {1\over N+1} \right)
\quad \quad \quad N = 1, 2, \ldots
\end{equation}
For $N=1, 3, 5,\ldots$, these singular points coincide with the ones 
in (\ref{4.1}).
However, there are additional divergencies of 
$c_V^{{\rm ex}}/k_B$ for $N = 2, 4, 6, \ldots$ when 
$c_V^{{\rm ex}}/k_B \to -\infty$, which is an unacceptable 
thermodynamic behaviour.
Therefore, in 2D, the analytic extension of the formula for
$c_V^{{\rm ex}}/k_B$ is meaningless, what supports the
arguments of Fisher et al \cite{Fisher} indicating the total
absence of any intermediate phase at nonzero particle density.

The situation is different in 1D.
The dominant configuration integral of the $z$-series at the
$2N$th order 
\begin{equation} \label{4.4}
{1\over L} \int_{\sigma}^L \prod_{i=1}^N {\rm d}x_i {\rm d}x'_i
{\prod_{i<j} \vert x_i - x_j \vert^{\beta}
\vert x'_i - x'_j \vert^{\beta}
\over \prod_{i,j} \vert x_i - x'_j \vert^{\beta}}
\end{equation} 
diverges for the line length $L\to\infty$ when $\beta<\beta_N$ with
\begin{equation} \label{4.5}
1D: \quad \quad \quad \beta_N = 2 \left( 1 - {1\over 2N}
\right), \quad \quad \quad N = 1, 2,\ldots
\end{equation}
The singularities of $c_V^{{\rm ex}}/k_B$, given by formula
(\ref{3.18}), originate in the region $1<\beta<2$ exclusively
from the term $\psi'((1-\beta)/(2-\beta))$.
According to (\ref{3.19}), $\psi'(x)$ has second-order poles
at $x=1-N$ $(N=1, 2,\ldots)$.
The corresponding ${\bar \beta}_N$ coincide just with $\beta_N$, 
so the analytic continuation of (\ref{3.18}) into the collapse region 
$1<\beta<2$ reproduces exactly the singularities suggested
by Gallavotti and Nicol{\'o}.
Moreover, $c_V^{{\rm ex}}/k_B$ is always {\it positive} in the
underlying region $1<\beta<2$.
It is therefore tempting to conjecture that the multipole
intermediate phases might exist in 1D.

The last remark concerns the generalization of the relation 
(\ref{3.11}) to the case of a small hard core $\sigma$:
\begin{equation} \label{4.6}
{n\over 2} = 2 \int_{\sigma}^{\lambda} {\rm d}x {z^2\over 
\vert x \vert^{\beta}} = {2 z^2 \over 1-\beta} 
\left( \lambda^{1-\beta} - \sigma^{1-\beta} \right)
\end{equation}
with $\lambda>\sigma$.
Then, around $\beta=1$, the free energy
\begin{equation} \label{4.7}
f \sim {1\over 2} \ln \left( {1-\beta \over \lambda^{1-\beta}
- \sigma^{1-\beta}} \right)
\end{equation}
is well defined for $\beta \to 1^-$ as well as $\beta \to 1^+$.
Applying (\ref{3.17b}) and taking the $\sigma\to 0$ limit,
one finds
\begin{equation} \label{4.8}
{c_V^{{\rm ex}} \over k_B} \sim {1\over 2(1-\beta)^2}
\quad \quad \quad {\rm for}\ {\rm both}\ \beta\to 1^-\ {\rm and}\ 
\beta\to 1^+
\end{equation}
i.e., the leading singular term of the expansion (\ref{3.20}) 
admits an analytic continuation from $\beta<1$ to the 
$\beta>1$ region.
This fact supports the above conjecture.

We hope to motivate numerical simulations of the model under
consideration.

Finding an integrable TC Coulomb gas with some short-range 
regularization of the interaction Coulomb potential 
might be a realistic task in one dimension. 

\renewcommand{\theequation}{A.\arabic{equation}}
\setcounter{equation}{0}

\section*{Appendix}
We consider the 1D TC log-gas confined to a straight line 
of size $L\to\infty$.
A pair of ($\pm$)-charged particles $i,j$ interacts via
\begin{equation} \label{A.1}
v(x_i,q_i \vert x_j,q_j)  =  q_i q_j v(x_i,x_j)
\end{equation}
where the Coulomb potential $v(x,x')$ is defined by (\ref{1.3}).
The supposed equality of the species fugacities $z_+=z_-=z$
corresponds to a neutral system with homogeneous particle densities
$n_+=n_-=n/2$.

The renormalized Mayer expansion in density (for details, see refs.
\cite{Samaj}, \cite{Deutsch}, \cite{Jancovici2}) is based on 
the expansion of each Mayer function in the inverse temperature $\beta$, 
and on the consequent series elimination of two-coordinated field
circles between every couple of three- or more-coordinated field
circles; by coordination of a circle we mean its bond coordination.
The renormalized bonds are given by
\begin{equation} \label{A.2}
{\begin{picture}(55,20)(0,7)
    \Photon(0,10)(55,10){1}{7}
    \BCirc(0,10){2.5} \BCirc(55,10){2.5}
    \Text(0,0)[]{$x,q$} \Text(55,0)[]{$x',q'$}
    \Text(28,23)[]{$K$} 
\end{picture}}
\ \ \ = \ \ \
{\begin{picture}(55,20)(0,7)
    \DashLine(0,10)(55,10){7}
    \BCirc(0,10){2.5} \BCirc(55,10){2.5}
    \Text(0,0)[]{$x,q$} \Text(55,0)[]{$x',q'$}
    \Text(28,23)[]{$-\beta v$} 
\end{picture}}\ \ \ +\ \ \
{\begin{picture}(110,20)(0,7)
    \DashLine(0,10)(55,10){5}
    \DashLine(55,10)(110,10){5}
    \BCirc(0,10){2.5} \BCirc(110,10){2.5}
    \Vertex(55,10){2.2}
    \Text(0,0)[]{$x,q$} \Text(110,0)[]{$x',q'$}
\end{picture}}\ \ \ + \ \ldots
\end{equation}
where, besides the integration over spatial coordinate $x$ of a field
(black) $n_q(x)$-circle, the summation over charge $q$-states 
at this vertex is assumed as well.
For the interaction under consideration (\ref{A.1}), the renormalized
bonds exhibit the same charge-dependence,
\begin{subequations}
\begin{equation} \label{A.3a}
K(x,q \vert x',q') = q q' K(x,x')
\end{equation}
where $K(x,x')=K(\vert x-x' \vert)$ is given by
\begin{equation} \label{A.3b}
K(x) = - {\beta\over 2} \int_{-\infty}^{\infty} \rd k
{\exp({\rm i} k x) \over \vert k \vert + \pi n \beta}
\end{equation}
\end{subequations}
By a simple rescaling of integration variable $k$, $K$ is shown 
to exhibit a special scaling form
\begin{subequations} \label{A.4} 
\begin{eqnarray} 
K(x) & = & - \beta {\bar K}(\pi n \beta \vert x \vert )
\label{A.4a} \\
{\bar K}(x) & = & {1\over 2} \int_{-\infty}^{\infty} \rd k
{\exp({\rm i} k x) \over \vert k \vert + 1} \label{A.4b}
\end{eqnarray}
\end{subequations}
The small-$x$ expansion of $K$ reads \cite{Gradshteyn}
\begin{equation} \label{A.5}
K(x) \sim \beta \left[ \ln(\pi n \beta \vert x \vert) + C
- {1\over 2} \pi^2 n \beta \vert x \vert + \ldots \right]
\end{equation}
where $C$ is the Euler's constant.
At asymptotically large-$x$ distances, by using twice integration per
partes it can be shown that $K$ decays algebraically as follows
\begin{equation} \label{A.6}
K(x) \sim - {\beta \over (\pi n \beta x)^2} 
\end{equation}
which exhibits the poor screening properties of log-gases in 1D
\cite{Forrester1}, \cite{Alastuey}.

The procedure of bond-renormalization transforms the ordinary Mayer 
representation of the dimensionless (minus) excess Helmholtz 
free energy, denoted as $\Delta[n]$, into
\begin{subequations} \label{A.7}
\begin{equation} \label{A.7a}
\Delta[n] = \ \ 
\begin{picture}(50,20)(0,7)
    \DashLine(0,10)(40,10){5}
    \Vertex(0,10){2.2} \Vertex(40,10){2.2}
\end{picture}
 + \ D^{(0)}[n] + \ \sum_{s=1}^{\infty}\ D^{(s)}[n] ,
\end{equation}
where 
\begin{equation} \label{A.7b}
D^{(0)}  =  \ \
\begin{picture}(40,20)(0,7)
    \DashCArc(20,-10)(28,45,135){5}
    \DashCArc(20,30)(28,225,315){5}
    \Vertex(0,10){2} \Vertex(40,10){2}
\end{picture} \ \ +\ \ 
\begin{picture}(40,20)(0,19)
    \DashLine(0,10)(40,10){5}
    \DashLine(0,10)(20,37){5}
    \DashLine(20,37)(40,10){5}
    \Vertex(0,10){2} \Vertex(40,10){2} \Vertex(20,37){2}
\end{picture} \ \ +\ \ 
\begin{picture}(30,20)(0,10)
    \DashLine(0,0)(30,0){5}
    \DashLine(0,0)(0,30){5}
    \DashLine(0,30)(30,30){5}
    \DashLine(30,0)(30,30){5}
    \Vertex(0,0){2} \Vertex(30,0){2} \Vertex(0,30){2} \Vertex(30,30){2}
\end{picture} \ \ +\ \ \ldots 
\end{equation}
\end{subequations}
is the sum of all unrenormalized ring diagrams and $\{ D^{(s)} \}$
represents the set of all remaining completely renormalized graphs;
the first few $D^{(s)}$-graphs are drawn in the sketch (11) of ref.
\cite{Samaj}.
The first term on the rhs of (\ref{A.7a}) is fixed to zero by c
harge neutrality.
The second term $D^{(0)}$, eq. (\ref{A.7b}), is expressible
as follows \cite{Samaj}
\begin{equation} \label{A.8}
D^{(0)} = {L\over 2} \int_0^n {\rm d} n' \lim_{x\to 0}
\left[ K(x,n') + \beta v(x) \right]
\end{equation}
With respect to (\ref{A.5}), one gets
\begin{equation} \label{A.9}
{D^{(0)} \over L} = {\beta \over 2} ( n \ln n - n) +
{\beta n \over 2} \left[ C + \ln (\pi \beta) \right]
\end{equation}
As concerns the completely renormalized graphs, due to 
the $\pm$ charge symmetry, only those $D^{(s)}$ are
nonzero whose all vertices have an even bond coordination.
The scaling form of the dressed bond $K$, formula (\ref{A.4}),
permits us to perform the $n$- and $\beta$-classification
of a nonzero diagram $D^{(s)}$, composed of $N_s$ vertices
(each bringing the factor $n$) and $L_s$ bonds.
Each dressed bond contributes by the factor $-\beta$ and
enforces the substitution $x' = x \pi n \beta$ which 
manifests itself as the factor $1/(\pi n \beta)$ for each
field-circle integration $\int {\rm d}x$.
Since there are $(N_s -1)$ independent field-circle
integrations in $D^{(s)}$, one concludes that
\begin{subequations}
\begin{equation} \label{A.10a}
{D^{(s)} \over L} = n \beta^{L_s-N_s+1} d_s
\end{equation}
where $d_s$ is the number
\begin{equation} \label{A.10b}
d_s = {D^{(s)}(n=1,\beta=1) \over L}
\end{equation}
\end{subequations}

The first nonzero diagram from the sketch (11) of ref. \cite{Samaj} is
\begin{subequations}
\begin{equation} \label{A.11a}
D^{(2)} = \ \ 
\begin{picture}(60,40)(0,7)
    \PhotonArc(20,-10)(28,45,135){1}{9}
    \PhotonArc(20,30)(28,225,315){1}{9}
    \PhotonArc(20,6)(20,15,165){1}{11}
    \PhotonArc(20,14)(20,195,345){1}{11}
    \Vertex(0,10){2} \Vertex(40,10){2}
\end{picture}
\end{equation}
It contributes to the $\beta^3$ order, with
\begin{equation} \label{A.11b}
d_2 = {1\over 2! 4!} \int_{-\infty}^{\infty} {{\rm d}x \over \pi}
{\bar K}^4(x)
\end{equation}
\end{subequations}
In the next $\beta^4$ order, only diagram 
\begin{subequations}
\begin{equation} \label{A.12a}
D^{(6)} = \ \ 
\begin{picture}(60,40)(0,16)
    \PhotonArc(32,6)(32,115,170){1}{7.5}
    \PhotonArc(-12,40)(32,295,355){1}{7.5}
    \PhotonArc(9,5)(32,10,70){1}{7.5}
    \PhotonArc(52,40)(32,185,250){1}{7.5}
    \PhotonArc(20,-17)(32,55,125){1}{7.5}
    \PhotonArc(20,39)(34.5,235,310){1}{9}
    \Vertex(0,10){2} \Vertex(40,10){2} \Vertex(20,35){2}
\end{picture}
\end{equation}
survives, and
\begin{equation} \label{A.12b}
d_6 = {1\over 3! (2!)^3} \int_{-\infty}^{\infty} {{\rm d}x_1 \over \pi}
\int_{-\infty}^{\infty} {{\rm d}x_2 \over \pi}
{\bar K}^2(x_1) {\bar K}^2(x_2) {\bar K}^2(\vert x_1-x_2 \vert))
\end{equation}
\end{subequations}
etc.
To evaluate $d_2$ and $d_6$, we Fourier-transform ${\bar K}^2(x)$:
\begin{eqnarray} \label{A.13}
{\hat G}(k) & = & \int_{-\infty}^{\infty} {{\rm d}x \over \sqrt{2\pi}}
{\rm e}^{-{\rm i}k x} ~ {\bar K}^2(x) \nonumber \\
& = & \sqrt{2\pi} {(1+\vert k \vert) \over \vert k \vert 
(2+\vert k\vert ) } \ln (1 +\vert k \vert )
\end{eqnarray}
In terms of ${\hat G}(k)$,
\begin{subequations}
\begin{eqnarray}
d_2 & = & {1\over 2! 4! \pi} \int_{-\infty}^{\infty} {\rm d}k ~
{\hat G}^2(k) \label{A.14a} \\
d_6 & = & {1\over 3! (2!)^3 \pi^2} \sqrt{2\pi} 
\int_{-\infty}^{\infty} {\rm d}k ~ {\hat G}^3(k) \label{A.14b}
\end{eqnarray}
\end{subequations}
With the aid of the substitution $1+k=\exp(t)$ ($k$ and $t$ positive),
\begin{subequations}
\begin{eqnarray} 
d_2 & = & {1\over 12}{1\over 2^3} \left[ \pi^2 + 7 \zeta(3) \right] 
\label{A.15a} \\
d_6 & = & {1\over 12}{1\over 2^4} \left[ \pi^2 + 6 \zeta(3) \right] 
\label{A.15b}
\end{eqnarray}
\end{subequations}
where $\zeta$ is the Riemann's zeta function.

Finally, for $q=\pm$,
\begin{equation} \label{A.16}
\ln \left( {n_q \over z} \right) = \ln \left( {n \over 2 z} \right) =
{\partial \over \partial n} \left[ {\Delta(n) \over L} \right]
\end{equation}
Using results of the previous paragraph, one ends up with
\begin{eqnarray} \label{A.17}
{n^{1-\beta/2}\over z} & = & 2 \beta^{\beta/2} \exp \left\{
(C + \ln \pi){\beta \over 2} + \sum_{s=1}^{\infty} d_s
\beta^{L_s-N_s+1} \right\} \nonumber \\
& = & 2 \beta^{\beta/2} \exp \big\{ (C+\ln \pi) {\beta\over 2} + 
{1\over 12} \left[ \pi^2 +7 \zeta(3) \right] \left( {\beta\over 2}
\right)^3  \nonumber \\
& & + {1\over 12} \left[ \pi^2 + 6 \zeta(3) \right] 
\left( {\beta\over 2} \right)^4  + O(\beta^5) \big\}
\end{eqnarray}
in agreement with (\ref{3.8}).

\section*{Acknowledgments}
I am grateful to Bernard Jancovici for stimulating discussions,
comments and careful reading of the manuscript.
My stay in LPT Orsay is supported by a NATO fellowship.
A partial support by Grant VEGA 2/7174/20 is acknowledged.

\end{document}